\documentclass{article}

{}
{}
{}

\usepackage{amsmath} 
\usepackage{amsthm} 
\usepackage{cite} 
\usepackage{hyperref} 
\usepackage{graphicx}
\usepackage{algorithmic} 
\usepackage{hyperref}
\usepackage{authblk}
\usepackage[margin=1.2cm]{geometry}
\usepackage{multicol}
\usepackage{blindtext}
\usepackage{epstopdf}
\usepackage{color}
\usepackage{float}
\usepackage{multirow}
\usepackage{amssymb}
\usepackage[ruled,linesnumbered]{algorithm2e}
\allowdisplaybreaks[4]

\begin{document}

\title{Robust Parametric Microgrid Dispatch Under Endogenous Uncertainty of Operation- and Temperature-Dependent Battery Degradation}

\author[1,2]{Rui Xie}
\author[1]{Jun Wang}
\author[1]{Jiaxu Duan}
\author[1]{Chao Ma}
\author[1]{Yunhui Liu}
\author[2*]{Yue Chen}

\affil[1]{State Key Laboratory of Environmental Adaptability for Industrial Products, China National Electric Apparatus Research Institute Co., Ltd., Guangzhou, China.}
\affil[2]{{Department of Mechanical and Automation Engineering, The Chinese University of Hong Kong, Hong Kong, China.}\newline *Email: yuechen@mae.cuhk.edu.hk}
\date{}

\setcounter{Maxaffil}{0}
\renewcommand\Affilfont{\itshape\small}

\maketitle

\begin{abstract}
Batteries play a critical role in microgrid energy management by ensuring power balance, enhancing renewable utilization, and reducing operational costs. However, battery degradation poses a significant challenge, particularly under extreme temperatures. This paper investigates the optimal trade-off between battery degradation and operational costs in microgrid dispatch to find a robust cost-effective strategy from a full life-cycle perspective. A key challenge arises from the endogenous uncertainty (or decision-dependent uncertainty, DDU) of battery degradation: Dispatch decisions influence the probability distribution of battery degradation, while in turn degradation changes battery operation model and thus affects dispatch. In this paper, we first develop an XGBoost-based probabilistic degradation model trained on experimental data across varying temperature conditions. We then formulate a parametric model predictive control (MPC) framework for microgrid dispatch, where the weight parameters of the battery degradation penalty terms are tuned through long-term simulation of degradation and dispatch interactions. Case studies validate the effectiveness of the proposed approach.
\end{abstract}

\begin{changemargin}{0.86cm}{0.86cm} 
\end{changemargin}

\begin{multicols}{2}

\section{Introduction}

The urgent need to reduce carbon emissions and promote green energy worldwide has driven the widespread use of photovoltaic (PV) power generation in microgrids. Since PV power is highly volatile and uncertain, energy storage system (ESS) has been combined with PV, forming the PV-ESS integrated system. In this context, ESS is critical to the reliable and economical operation of microgrids, which helps to maintain power balance, improve the utilization of intermittent PV sources, and reduce operational costs.

The ESS is managed and operated through microgrid dispatch strategies. Research on renewable-integrated microgrid dispatch has mostly focused on handling renewable and load uncertainties to prevent supply-demand imbalances \cite{yang2025robust}. However, this focus overlooks a critical consequence: the frequent charging and discharging cycles required to smooth renewable intermittency accelerate battery degradation. This problem becomes more serious under extreme temperature conditions \cite{liu2025degradation}, leading to a shorter operational life and a significant increase in life cycle costs. Therefore, incorporating battery degradation into the dispatch optimization problem is essential for the long-term economics of microgrids.

To this end, researchers have developed various battery degradation models for integration into power system optimization \cite{xu2022role}. Some approaches represented degradation with simplified linear models or as a basic penalty in the objective function \cite{bahloul2024optimal}. Recognizing that degradation is a complex phenomenon, multi-stress models were introduced \cite{wang2020impact}. For instance, semi-empirical models considering the combined impacts of state of charge (SOC), depth of discharge (DOD), and equivalent full cycle (EFC) numbers were formulated and integrated into microgrid scheduling \cite{wang2024optimal}. To improve tractability within optimization frameworks, these multi-factor degradation dynamics have also been expressed as mixed-integer linear programming constraints \cite{liu2023MILP}.

Recently, the development of machine learning has promoted the emergence of powerful data-driven degradation models. A neural network trained on simulation data was used in day-ahead microgrid scheduling via a heuristic algorithm \cite{zhao2024microgrid}. Similarly, an extreme gradient boosting (XGBoost) model was employed to predict the remaining useful life of batteries within an energy management system for an agricultural microgrid \cite{safavi2025battery}.

Despite these advancements, significant challenges still exist in the literature. First, while most studies penalize degradation in the dispatch objective function, quantifying the penalty coefficient or the equivalent cost of degradation remains a critical issue. This coefficient determines the trade-off between immediate operational costs and long-term battery degradation, and an arbitrary choice can lead to suboptimal life-cycle performance. Second, the majority of degradation models employed in optimization are deterministic, which contradicts the stochastic nature of the degradation process observed in experiments \cite{chen2024lithium}. This leads to strategies that are not robust enough facing real-world uncertainty.

Third, existing studies fail to adequately model the dynamic and bidirectional interactions between dispatch decisions and battery degradation. Dispatch decisions influence the rate of degradation; concurrently, the accumulated degradation constrains the operational capabilities of the ESS in subsequent dispatch periods. This mutual impact is an example of endogenous uncertainty (or decision-dependent uncertainty, DDU) \cite{tan2025adjustable}, where the probability distribution of a random variable is affected by decisions made before. This persistent interaction requires a multi-stage framework with DDU to model, whose general solution methods are lacking.

To address these research gaps, this paper proposes a novel robust parametric microgrid dispatch framework that explicitly accounts for the DDU of battery degradation. Our objective is to determine the optimal life-cycle trade-off between operational costs and battery degradation. The main contributions of this work are twofold:

1) \emph{Probabilistic degradation model:} We develop an XGBoost-based probabilistic model trained on experimental data from 196 lithium-ion batteries \cite{wildfeuer2023experimental}. This model captures the uncertain nature of degradation by predicting quantiles of the capacity degradation rate as a function of key stress factors, including temperature, DOD, current capacity, and maximum charging/discharging power.

2) \emph{Robust parametric dispatch framework for degradation-related DDU:} We formulate a model predictive control (MPC)-based dispatch model that incorporates tunable weight parameters for multiple degradation-related penalty terms. A robust optimization (RO) problem with DDU is formulated to optimize these parameters over a long-term horizon, considering the interactions between microgrid dispatch and battery degradation. This problem is solved using a simulation-based particle swarm optimization (PSO) approach.

The remainder of this paper is organized as follows: Section 2 establishes the probabilistic battery degradation model. Section 3 introduces the optimal microgrid dispatch method considering battery degradation. Section 4 presents case study results, and Section 5 concludes the paper.

\section{Probabilistic Battery Degradation Model Using Experimental Data}

In this section, we first introduce the experimental data processing and then develop the degradation model.

\subsection{Battery Degradation Data Processing}
\label{sec:data-process}

The experimental data for the battery degradation model are from \cite{wildfeuer2023experimental}, which were collected during the calendar aging or cyclic aging of 196 lithium-ion batteries (Sony/Murata US18650VTC5) at different ambient temperatures. Check-up (CU) procedures were performed periodically to characterize the state of health. 

We process the detailed voltage and current records from these experiments to quantify performance indicators and operational stress factors. The following metrics are calculated for the interval between each pair of consecutive CU procedures:
\begin{subequations}
\label{eq:measure}
\begin{align}
    \label{eq:measure-C}
    & C = \sqrt{e^{C, CU} e^{D, CU}}, \\
    \label{eq:measure-DOD}
    & DOD = \sqrt{e^{C, CYC} e^{D, CYC}} / C_0, \\
    \label{eq:measure-p}
    & \overline{p}^C = \max_t p_t^{C, CYC},\; \overline{p}^D = \max_t p_t^{D, CYC}. 
\end{align}
\end{subequations}
The energy capacity $C$ is computed in \eqref{eq:measure-C} as the geometric mean of the accumulated charging/discharging energy ($e^{C, CU}$ and $e^{D, CU}$) during a constant current constant voltage charging and discharging cycle in the CU procedure. Equations \eqref{eq:measure-DOD} and \eqref{eq:measure-p} are for cyclic aging cases. In \eqref{eq:measure-DOD}, the DOD value is the ratio of the geometric mean of the charging/discharging energy ($e^{C, CYC}$ and $e^{D, CYC}$) in a cycle to the initial energy capacity $C_0$. The maximum charging/discharging power ($\overline{p}^C$ and $\overline{p}^D$) are defined in \eqref{eq:measure-p} using the power values in a cycle, where $t$ is the time index.

With these metrics, we then calculate the target variable for our model: the capacity degradation rate. This rate is determined differently for calendar and cyclic aging.

For intervals of pure calendar aging, the degradation rate $r_i^{CAL}$ between the $i$-th CU and the $(i + 1)$-th CU is defined as the average daily capacity degradation:
\begin{align}
    r_i^{CAL} = (C_i - C_{i + 1}) / N_i^{DAY},
\end{align}
where $C_i$ and $C_{i+1}$ are the capacities measured at two consecutive CUs, and $N_i^{DAY}$ is the number of days between them. We utilize experimental data from different interval lengths and computationally eliminate the additional degradation induced by the CU procedure \cite{wildfeuer2023experimental}.

For cyclic aging experiments, the degradation rate $r_i^{CYC}$ is normalized by the EFC number:
\begin{subequations}
\begin{align}
    & r_i^{CYC} = (C_i - C_{i + 1}) / EFC_i, \\
    & EFC_i = DOD_i \cdot N_i^{CYC},
\end{align}
\end{subequations}
where $r_i^{CYC}$ represents the capacity degradation per EFC. The total EFC number for an interval is the product of the DOD and the number of cycles performed, $N_i^{CYC}$. 

The above data processing yields a structured dataset where each sample corresponds to the aging interval between two consecutive CUs. Each sample contains a set of features (capacity at the start of the interval, ambient temperature, DOD, and maximum charging/discharging power) and a corresponding label (the measured capacity degradation rate). This dataset will be used for training the probabilistic degradation model.

\subsection{XGBoost-Based Probabilistic Battery Degradation Model}

We develop a probabilistic model to characterize the probability distribution of the degradation rate. We utilize XGBoost \cite{chen2016xgboost}, which is a powerful and efficient implementation of tree boosting system, known for its ability to capture complex nonlinear relationships in data while mitigating overfitting.

To achieve a probabilistic forecast for the degradation rate $r$, we employ quantile regression rather than predicting a single value. For a given quantile $q \in (0, 1)$, the model is expected to predict a value $R_q$ such that the probability $\Pr [ r \leq R_q ]$ is $q$. This is achieved within the XGBoost framework by using the quantile loss function \cite{takeuchi2006nonparametric}. For a target quantile $q$, the loss for a true value $r$ and a predicted quantile $\hat{r}_q$ is defined as:
\begin{equation}
\label{eq:quantile_loss}
L_q(r, \hat{r}_q) =
\begin{cases}
q(r - \hat{r}_q), & \text{if}~ r \geq \hat{r}_q, \\
(1-q)(\hat{r}_q - r), & \text{if}~ r < \hat{r}_q.
\end{cases}
\end{equation}
This loss function asymmetrically penalizes overestimation and underestimation. For example, when predicting the $0.9$-quantile, the penalty for an underestimation ($\hat{r}_q < r$) is $9$ times higher than an overestimation ($\hat{r}_q > r$), forcing the model to learn the upper tail of the distribution. When $q=0.5$, the model is trained to predict the median.

For a chosen set of quantiles $Q = \{q_1, q_2, \dots, q_K\}$ with $0 < q_1 < q_2 < \dots < q_K < 1$, we train a distinct XGBoost model for each $q_k \in Q$. The final probabilistic degradation model is therefore a function that maps a set of input features to a vector of quantile predictions for the degradation rate. For example, to predict the capacity degradation rate in cyclic aging, for an input vector $\boldsymbol{x} = (C, \mathcal{T}, DOD, \overline{p}^C, \overline{p}^D)$ comprising the current capacity $C$, temperature $\mathcal{T}$, DOD, and maximum charging/discharging power ($\overline{p}^C$ and $\overline{p}^D$), the model outputs a vector $\hat{\boldsymbol{r}} = (\hat{r}_{q_1}, \hat{r}_{q_2}, \dots, \hat{r}_{q_K})$ of predicted degradation rates for the specified quantiles. Moreover, the vector $\hat{\boldsymbol{r}}$ is sorted to ensure $\hat{r}_{q_1} \leq \hat{r}_{q_2} \leq \dots \leq \hat{r}_{q_K}$. The interval $[\hat{r}_{q_j}, \hat{r}_{q_k}]$ is a prediction interval of probability $q_k - q_j$ for any $q_j < q_k$. We usually assume $q_j + q_k = 1$ to make the quantiles symmetric about $0.5$. These prediction intervals quantify the DDU of battery degradation, which is essential for the optimal dispatch framework developed in the next section.

\emph{Remark 1:} The battery's internal temperature is influenced by charging and discharging processes. In our formulation, the input $\mathcal{T}$ represents the ambient temperature, which serves as a boundary condition. The XGBoost model is trained to learn the complex relationship between the operational inputs and the resulting degradation, which considers the charging and discharging power by $\overline{p}^C$ and $\overline{p}^D$. Therefore, the model implicitly captures the thermal consequences of operation, as these effects are embedded in the training data.

\emph{Remark 2:} A gap exists between single-battery and ESS degradation. However, this can be overcome by correcting cell-level data for system-level effects \cite{liu2019degradation}. Moreover, the proposed model can be extended to construct an ESS degradation model if experiments are performed on the ESS.

\section{Optimal Microgrid Dispatch Considering Battery Degradation}

In this section, we first introduce the PV-ESS integrated microgrid, followed by the formulation of the parametric microgrid dispatch problem. We then formulate the parameter optimization as a RO problem and finally present the solution method.

\subsection{PV-ESS Integrated Microgrid}

\begin{figure}[H]
\centering
\includegraphics[width=0.8\linewidth]{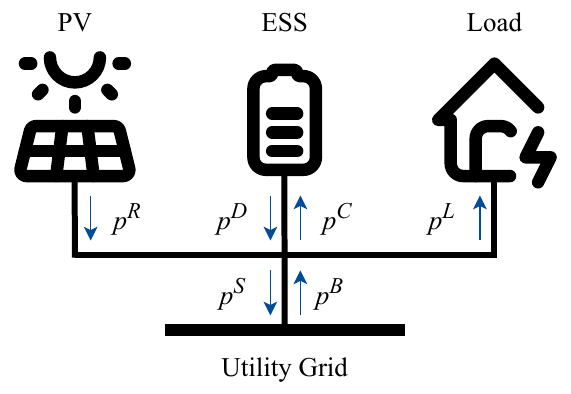}
\caption{Schematic of the PV-ESS integrated microgrid and power flow definitions.}
\label{fig:microgrid}
\end{figure}

The system under consideration is a grid-connected microgrid that integrates PV, ESS, and local electrical loads. A schematic of this architecture is presented in Figure~\ref{fig:microgrid}. The power sources include PV generation ($p^R$), discharging power of the ESS ($p^D$), and power purchased from the utility grid ($p^B$). The system's power consumption includes the load demand ($p^L$), charging power of the ESS ($p^C$), and power sold back to the utility grid ($p^S$).

Within this microgrid, we need to manage two classes of uncertainty: 1) Decision-independent uncertainty (DIU): The PV power generation and the load demand are instances of DIU, which are driven by external factors such as weather conditions and consumer behaviors, and are not influenced by the microgrid's dispatch decisions. 2) DDU: The degradation of the ESS is an endogenous uncertainty. The probability distribution of the battery's future capacity is affected by the historical dispatch decisions, particularly the charging and discharging power profiles.

\subsection{Parametric Microgrid Dispatch Considering Battery Degradation}

We formulate a parametric dispatch model, which optimizes the microgrid's operation over a finite time horizon $t = 1, 2, \dots, T$, balancing immediate operational costs with battery degradation penalty terms as follows:
\begin{subequations}
\label{eq:mpc}
\begin{align}
    \label{eq:mpc-a}
    \min_{\substack{\underline{e}, \overline{e}, \overline{p}^C, \overline{p}^D, \\ \boldsymbol{p^B}, \boldsymbol{p^S}, \boldsymbol{p^C}, \boldsymbol{p^D}, \boldsymbol{e}}}~ & \sum_{t = 1}^T (\lambda_t^B p_t^B - \lambda_t^S p_t^S) \\
    \label{eq:mpc-b}
    & + \theta^{EFC} \sum_{t = 1}^T (p_t^C + p_t^D) \tau \\
    \label{eq:mpc-c}
    & + \theta^{DOD} (\overline{e} - \underline{e}) \\
    \label{eq:mpc-d}
    & + \theta^C \overline{p}^C + \theta^D \overline{p}^D \\
    \label{eq:mpc-e}
    \mbox{s.t.}~ & p_t^B, p_t^S, p_t^C, p_t^D \geq 0, \forall t, \\
    \label{eq:mpc-f}
    & p_t^B + p_t^D + \hat{p}_t^R = p_t^S + p_t^C + \hat{p}_t^L, \forall t, \\
    \label{eq:mpc-g}
    & e_t = e_{t - 1} + p_t^C \tau \eta^C - p_t^D \tau / \eta^D, \forall t, \\
    \label{eq:mpc-h}
    & \underline{e} \leq e_t \leq \overline{e}, \forall t, \\
    \label{eq:mpc-i}
    & e_T = e_0, \underline{E} \leq \underline{e} \leq \overline{e} \leq \overline{E}, \\
    \label{eq:mpc-j}
    & p_t^C \leq \overline{p}^C, p_t^D \leq \overline{p}^D, \forall t, \\
    \label{eq:mpc-k}
    & \overline{p}^C \leq \overline{P}^C, \overline{p}^D \leq \overline{P}^D, \\
    \label{eq:mpc-l}
    & p_t^B \leq P^U, p_t^S \leq P^U, \forall t.
\end{align}
\end{subequations}
In the objective function, the first term \eqref{eq:mpc-a} represents the operational cost, calculated as the net cost of purchasing power from the grid and selling power to it, where $\lambda_t^B$ and $\lambda_t^S$ are purchase and selling electricity prices, respectively. The subsequent terms \eqref{eq:mpc-b}–\eqref{eq:mpc-d} are the degradation penalty terms designed for the key stress factors: \eqref{eq:mpc-b} penalizes the total energy throughput, which serves as a proxy for the EFC; \eqref{eq:mpc-c} penalizes the operational range of the stored energy ($\overline{e} - \underline{e}$), which is a proxy for the DOD; and \eqref{eq:mpc-d} penalizes the maximum charging and discharging power values ($\overline{p}^C$ and $\overline{p}^D$). The trade-off between operational cost and battery degradation is governed by the nonnegative weight parameter vector $\boldsymbol{\theta} = (\theta^{EFC}, \theta^{DOD}, \theta^C, \theta^D)$. The determination of $\boldsymbol{\theta}$ will be detailed later in this section.

The optimization is subject to a set of operational constraints. Constraint \eqref{eq:mpc-e} defines nonnegative power variables. Constraint \eqref{eq:mpc-f} enforces the microgrid's power balance in each time step $t$, using forecasts for PV generation ($\hat{p}_t^R$) and load ($\hat{p}_t^L$). The ESS energy dynamics are modeled in \eqref{eq:mpc-g}, where $e_t$ is the stored energy, and $\eta^C$ and $\eta^D$ are the charging and discharging efficiencies. Constraints \eqref{eq:mpc-h} and \eqref{eq:mpc-i} restrict the ESS energy level within the operating range $[\underline{e}, \overline{e}]$, which is itself bounded by the battery's current physical capacity limits $[\underline{E}, \overline{E}]$. The constraint $e_T = e_0$ ensures operational continuity for the future. Constraints \eqref{eq:mpc-j} and \eqref{eq:mpc-k} enforce limits on the ESS power, bounded by the physical capacities $\overline{P}^C$ and $\overline{P}^D$. Finally, constraint \eqref{eq:mpc-l} defines the boundaries for the power flow at the grid connection point.

This optimization problem is solved within a rolling-horizon MPC framework to handle the DIUs of PV generation and load. At the start of each dispatch horizon, problem \eqref{eq:mpc} is solved using the latest forecasts. Only the dispatch decisions for the first time step are implemented. After the first time step, new forecasts become available, and the process is repeated. This adaptive approach allows the dispatch strategy to continuously correct for forecast errors and respond to real-time conditions.

However, deploying this parametric MPC framework still faces two critical challenges: 1) The choice of the parameter vector $\boldsymbol{\theta}$ determines the trade-off between short-term costs and long-term degradation, influencing the total life-cycle cost of the system. 2) The dispatch strategy determined by $\boldsymbol{\theta}$ causes battery degradation. This degradation in turn reduces the battery's capacity, meaning the physical bounds $\underline{E}$ and $\overline{E}$ shrink over time. Optimally tuning the parameters while accounting for the degradation they cause is the focus of the RO method presented next.

\subsection{Robust Parameter Optimization}

To select a parameter vector $\boldsymbol{\theta}$ for the dispatch model that performs best over the entire lifecycle, we formulate the problem as a RO problem that seeks to minimize the total life-cycle cost under worst-case degradation scenarios.

First, we formalize the link between dispatch decisions and future battery capacity. The capacity at the end of a relatively long operational period of length $\Gamma$ (e.g., several months) is the initial capacity minus the degradation from cyclic and calendar aging:
\begin{align}
    c(\boldsymbol{p}, C) := C - r^{CYC}(\boldsymbol{p}, C) \cdot EFC(\boldsymbol{p}) - r^{CAL} \cdot \Gamma.
\end{align}
Here, $\boldsymbol{p}$ represents the collection of dispatch decisions over the period, and $EFC(\boldsymbol{p})$ is the total EFC number. Since the degradation rates $r^{CYC}$ and $r^{CAL}$ are uncertain, so is $c(\boldsymbol{p}, C)$.

To ensure the strategy is robust, we use the probabilistic degradation model to determine the $q$-quantile forecasts $r_q^{CYC}$ and $r_q^{CAL}$ for the degradation rates, where $q \in (0, 1)$ is specified by the decision maker. A higher $q$ corresponds to a more conservative result. This allows us to define the worst-case capacity as a function of the current state $C$ and the chosen dispatch policy $\boldsymbol{p}$:
\begin{align}
    \label{eq:capacity-degradation}
    h(\boldsymbol{p}, C) := C - r_q^{CYC}(\boldsymbol{p}, C) \cdot EFC(\boldsymbol{p}) - r_q^{CAL} \cdot \Gamma.
\end{align}
The uncertainty set for the next capacity state is defined as $\mathcal{U}(\boldsymbol{p}, C) := [h(\boldsymbol{p}, C), +\infty)$.

Then we formulate the RO problem in \eqref{eq:param}. The goal is to find the optimal parameter $\boldsymbol{\theta}$ and the resulting battery lifetime $N$ that minimize the total present value of costs over a project lifetime of $N_0$ periods.
\begin{subequations}
\label{eq:param}
\begin{align}
    & \min_{\substack{\boldsymbol{\theta} \geq 0, N_0 \geq N \in \mathbb{N}_+, \\ \boldsymbol{p_1} = \boldsymbol{f_1}(C_0, \boldsymbol{\theta})}} \Bigg(I_0 + \frac{g(\boldsymbol{p_1})}{1 + i} + \max_{\substack{C_1 \in \mathcal{U}(\boldsymbol{p_1}, C_0), \\ \boldsymbol{p_2} = \boldsymbol{f_2}(C_1, \boldsymbol{\theta})}} \frac{g(\boldsymbol{p_2})}{(1 + i)^2} \nonumber \\
    \label{eq:param-1}
    & + \dots + \max_{\substack{C_{N - 1} \in \mathcal{U}(\boldsymbol{p_{N - 1}}, C_{N - 2}), \\ \boldsymbol{p_N} = \boldsymbol{f_N}(C_{N - 1}, \boldsymbol{\theta})}} \frac{g(\boldsymbol{p_N})}{(1 + i)^N} \Bigg) \cdot \frac{1 - \frac{1}{(1 + i)^{N_0}}}{1 - \frac{1}{(1 + i)^N}} \\
    \label{eq:param-2}
    & \mbox{s.t.}~ C_n \geq \underline{C}, \forall C_n \in \mathcal{U}(\boldsymbol{p_n}, C_{n - 1}), n = 1, 2, \dots, N - 1.
\end{align}
\end{subequations}
The objective is to minimize the total cost. This includes the initial investment in the ESS $I_0$, plus the sum of discounted operational costs $g(\boldsymbol{p}_n)$ over the battery's life $N$, where $i$ is the interest rate. The battery is considered retired when its capacity falls below the threshold $\underline{C}$, and a new ESS will be installed to replace it. The total present value of this replacement chain is calculated by multiplying a normalization coefficient.

The operation strategy determined by problem \eqref{eq:mpc} in the MPC framework is denoted by $\boldsymbol{p_n} = \boldsymbol{f_n}(C_{n - 1}, \boldsymbol{\theta})$ for $n = 1, 2, \dots, N$, where the function notation $\boldsymbol{f_n}(C_{n - 1}, \boldsymbol{\theta})$ emphasizes its dependence on the current capacity $C_{n - 1}$, penalty term parameter $\boldsymbol{\theta}$, and the PV and load uncertainties in that operation period. The capacity $C_{n - 1}$ appears in \eqref{eq:mpc} via $\underline{E} = (C_0 - C_{n - 1})/2$ and $\overline{E} = (C_0 + C_{n - 1})/2$. 

We employ the RO technique to ensure robustness using the ``max'' operators and considering the worst-case battery degradation in the uncertainty set $\mathcal{U}(\boldsymbol{p}, C)$. Moreover, since the ranges of the uncertainty sets depend on the first-stage decision $\boldsymbol{\theta}$, problem \eqref{eq:param} is a RO problem incorporating the DDU of battery degradation.

\subsection{Solution Method}

The RO problem \eqref{eq:param} is computationally intractable to solve directly due to its nested ``max'' operators and the dependency of the dispatch strategy $\boldsymbol{p}_n$ on the parameter vector $\boldsymbol{\theta}$. However, we can exploit the structure of the problem to develop an efficient solution method.

The objective is to find a policy $\boldsymbol{\theta}$ that minimizes the total life-cycle cost. We claim that the total cost from now on is decreasing in the battery's current capacity, which has two reasons: 1) A lower capacity means the battery will reach its end-of-life threshold $\underline{C}$ sooner. 2) A lower capacity restricts the battery's operational flexibility within the MPC dispatch problem \eqref{eq:mpc}. This reduced ability to charge and discharge leads to higher operational costs. Consequently, the ``max'' operator is resolved by always selecting the lower bound of the uncertainty set, i.e., $C_n = h(\boldsymbol{p}_n, C_{n-1}), \forall n$. This insight will also be verified in the case studies. It allows us to reformulate the RO problem \eqref{eq:param} into a deterministic optimization problem as follows:
\begin{subequations}
\label{eq:opt}
\begin{align}
    \label{eq:opt-1}
    \min_{\boldsymbol{\theta}, N, C, \boldsymbol{p}}~ & \frac{1 - \frac{1}{(1 + i)^{N_0}}}{1 - \frac{1}{(1 + i)^N}} \cdot \left( I_0 + \sum_{n = 1}^N \frac{g(\boldsymbol{p_n})}{(1 + i)^n} \right) \\
    \mbox{s.t.}~ & \boldsymbol{\theta} \geq 0, N_0 \geq N \in \mathbb{N}_+, \\
    & \boldsymbol{p_n} = \boldsymbol{f_n}(C_{n - 1}, \boldsymbol{\theta}), n = 1, 2, \dots, N, \\
    & \underline{C} \leq C_n = h(\boldsymbol{p_n}, C_{n - 1}), n = 1, 2, \dots, N. 
\end{align}
\end{subequations}
The total cost is a function of $\boldsymbol{\theta}$, but we cannot compute its gradient because it involves iteratively solving the MPC dispatch problem \eqref{eq:mpc}. This structure makes the problem suited for a derivative-free, simulation-based optimization approach. We employ a hybrid method where the PSO algorithm \cite{kennedy1995particle} forms the outer loop to search for the optimal parameter vector $\boldsymbol{\theta}$. The cost of each candidate vector is evaluated using a life-cycle simulation, as described in Algorithm~\ref{alg:simulation}. PSO does not guarantee convergence to the global optimum for this complex and nonconvex problem, but it can effectively search the solution space without requiring gradient information.

\begin{algorithm}[H]
\caption{Life-cycle simulation for a given $\boldsymbol{\theta}$}
\label{alg:simulation}
\DontPrintSemicolon
\SetAlgoLined
\SetKwInOut{Input}{Input}
\SetKwInOut{Output}{Output}
\Input{Parameter vector $\boldsymbol{\theta}$, initial capacity $C_0$, end-of-life capacity $\underline{C}$, project horizon $N_0$, interest rate $i$, investment cost $I_0$, quantile $q$}
\Output{Total life-cycle cost for $\boldsymbol{\theta}$}
\BlankLine
Initialize: $C \leftarrow C_0$, $n \leftarrow 0$;\;
\While{$n < N_0$ and $C \geq \underline{C}$}{
$n \leftarrow n + 1$;\;
Simulate MPC operation for the $n$-th period of length $\Gamma$ and obtain 
$\boldsymbol{p_n} \leftarrow \boldsymbol{f_n}(C, \boldsymbol{\theta})$;\;
Use the degradation model to predict $r_q^{CYC}(\boldsymbol{p_n}, C)$ and $r_q^{CAL}$;\;
$C \leftarrow h(\boldsymbol{p_n}, C)$;\;
}
$N \leftarrow n$;\;
Calculate the total life-cycle cost using \eqref{eq:opt-1};\;
\Return{Total life-cycle cost}
\end{algorithm}

\emph{Remark 3:} The total computation time $T_{total}$ of the proposed solution method can be approximated as:
\begin{align}
    T_{total} \approx N_{particles} \cdot N_{iter} \cdot \overline{N} \cdot \Gamma \cdot T_{MPC},
\end{align}
where $N_{particles}$ is the number of particles in the PSO algorithm, $N_{iter}$ is the number of PSO iterations, $\overline{N}$ is the average number of loops in Algorithm~\ref{alg:simulation} (linear in the lifetime of the battery), and $T_{MPC}$ is the average time required to solve the MPC dispatch problem \eqref{eq:mpc} for a single time step. The MPC dispatch can be solved efficiently since it is a small-scale linear program, with its complexity being polynomial in the length of the MPC horizon. Compared with conventional dispatch strategies, our method requires significantly more computational time to determine the parameter vector $\boldsymbol{\theta}$, but provides an optimized and robust policy that effectively balances operational cost and uncertain battery degradation.

\section{Case Study}

In this section, we validate the proposed framework and demonstrate its effectiveness through a detailed case study. First, we outline the experimental setup and key parameters. Second, we evaluate the performance of the probabilistic battery degradation model. Finally, we analyze the performance of the proposed microgrid dispatch method.

All simulations and optimizations were implemented in Python 3.12, utilizing Gurobi 11.0.3 for solving the linear programming dispatch problems. The computations were performed on a laptop with an Intel Core i5-1335U processor and 16 GB RAM.

\subsection{Settings}

The case study considers a PV-ESS integrated microgrid. The ESS has an initial capacity of $910.8$ kWh and an investment cost of \$$2 \times 10^5$. It will be retired when its capacity degrades to 40\% of its initial capacity. For the base case, we assume an ambient temperature of $35$ °C. The microgrid adopts a time-of-use electricity tariff, where the purchase prices range from \$0.12 to \$0.50 per kWh. The selling price is lower than the purchase price at any time. The interest rate is $3.2$\% per year. Complete details of the input data are available in \cite{xie2025github}. In the life-cycle simulation, the operation period length $\Gamma$ is a season, and each season is represented by a typical day with operational costs and battery degradation scaled up proportionally.

\subsection{Results of Battery Degradation Model}

We first evaluate the prediction performance and then analyze the stress factors of battery degradation using the trained model.

\subsubsection{Prediction Performance}

The experimental dataset from \cite{wildfeuer2023experimental} was processed as described in Section~\ref{sec:data-process}, yielding 2,537 data points. They were partitioned into a training set (1,521 samples), a validation set (507 samples) for hyperparameter tuning, and a test set (508 samples) for performance evaluation.

\begin{figure}[H]
\centering
\includegraphics[width=0.9\linewidth]{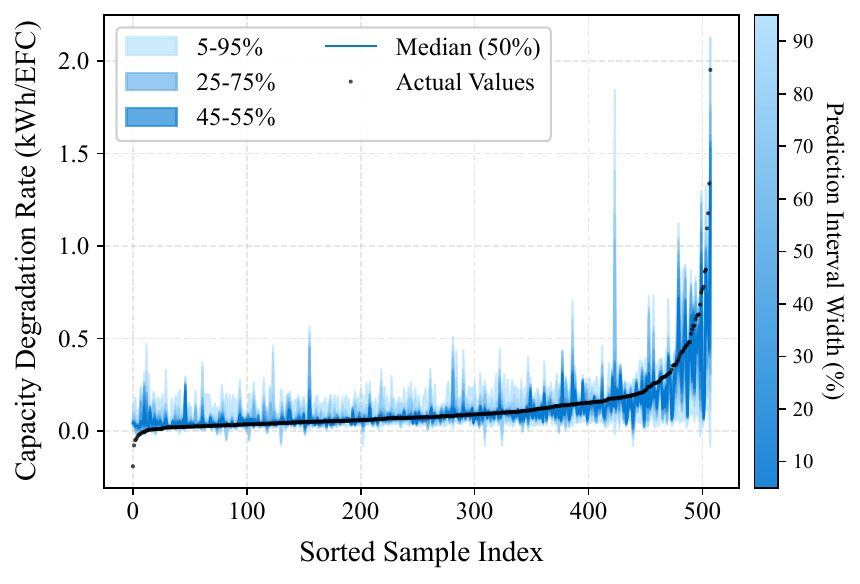}
\caption{Prediction intervals for samples in the test dataset, sorted according to the actual values.}
\label{fig:nested_quantiles}
\end{figure}

Figure~\ref{fig:nested_quantiles} provides a visualization of the model's probabilistic forecasting performance on the test data. The plot compares the model's prediction intervals against the actual measured degradation rates. The samples are sorted by their actual values. The figure demonstrates that the prediction intervals successfully envelop the vast majority of the actual values.

\begin{figure}[H]
\centering
\includegraphics[width=0.9\linewidth]{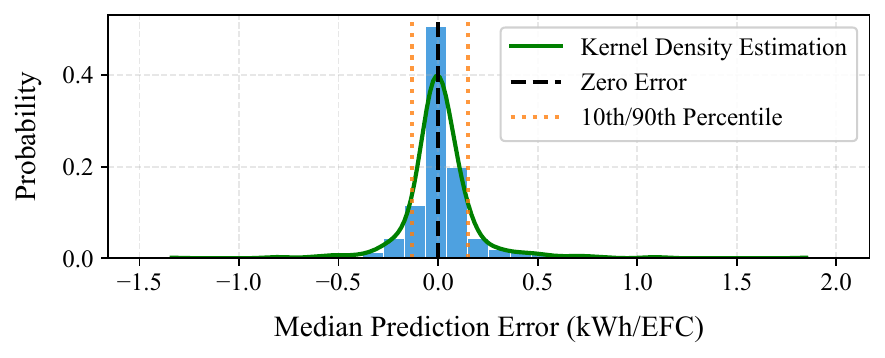}
\caption{Distribution of the median prediction error in the test dataset.}
\label{fig:error_distribution}
\end{figure}

Figure~\ref{fig:error_distribution} shows the distribution of the median prediction error (actual value minus $0.5$-quantile prediction). The distribution is tightly centered and symmetric around zero, indicating that the median prediction is unbiased. The 10th and 90th percentiles of the error are $-0.13$ kWh/EFC and $0.15$ kWh/EFC, respectively, showing that the magnitude of the prediction errors is small.

\subsubsection{Analysis of Degradation Stress Factors}

\begin{figure*}
\centering
\includegraphics[width=0.9\linewidth]{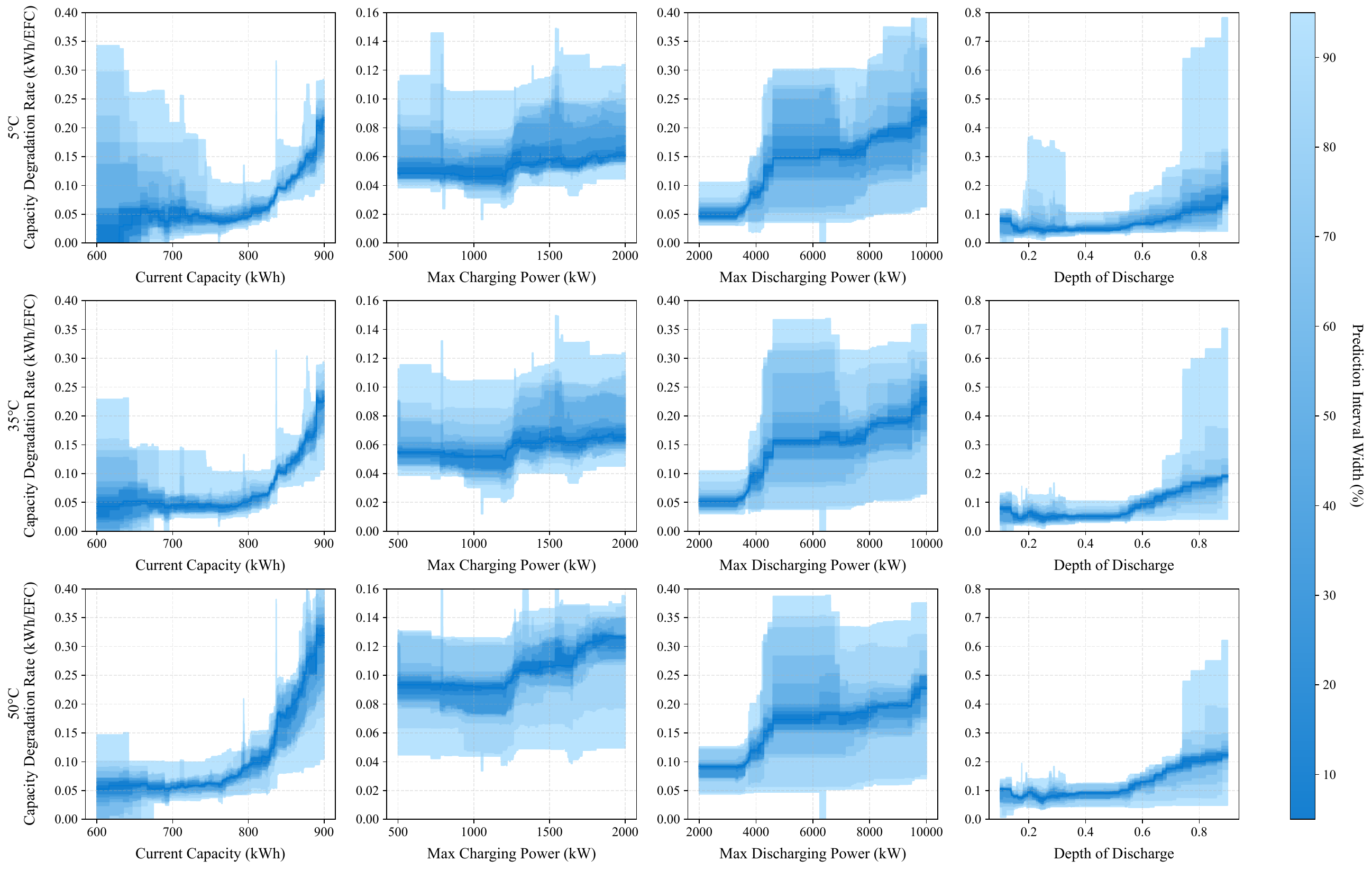}
\caption{Prediction intervals for capacity degradation rate under different settings.}
\label{fig:temperature_comparison}
\end{figure*}

We now use the trained XGBoost model to perform a sensitivity analysis for the impacts of the stress factors on the degradation rate distribution. In the base case, we use the following parameters: $C = 800$ kWh, $DOD = 40$\%, $\overline{p}^C = 1,000$ kW, and $\overline{p}^D = 2,000$ kW. We then vary these stress factors and ambient temperatures. The results are presented in Figure~\ref{fig:temperature_comparison}. As the results indicate, the battery ages differently at various ambient temperatures. For example, the median degradation rates are higher at temperature $\mathcal{T} = 50$ °C than the values at $\mathcal{T} = 35$ °C. In addition, the dispersion degree of the degradation uncertainty changes a lot under different settings, which shows the necessity of establishing a probabilistic and multi-stress prediction model. In addition, the results also verify that the battery degradation uncertainty is decision-dependent.

\subsection{Results of Microgrid Dispatch}

We first compare the RO method with two alternative strategies and then conduct a sensitivity analysis on key parameters.

\subsubsection{Method Comparison}

To demonstrate the value of our risk-averse RO approach, we compare its performance against two alternatives:
\begin{itemize}
    \item Benchmark: Battery degradation is ignored in the microgrid dispatch model \eqref{eq:mpc} by setting $\boldsymbol{\theta} = \boldsymbol{0}$.
    \item SP: A strategy that accounts for degradation but in a risk-neutral manner. It uses random forecasts from our probabilistic model, rather than a worst-case value.
    \item RO: The proposed RO method with $q = 0.90$.
    \item No Usage: Do not charge or discharge the ESS during microgrid operation.
\end{itemize}

\begin{figure}[H]
\centering
\includegraphics[width=0.9\linewidth]{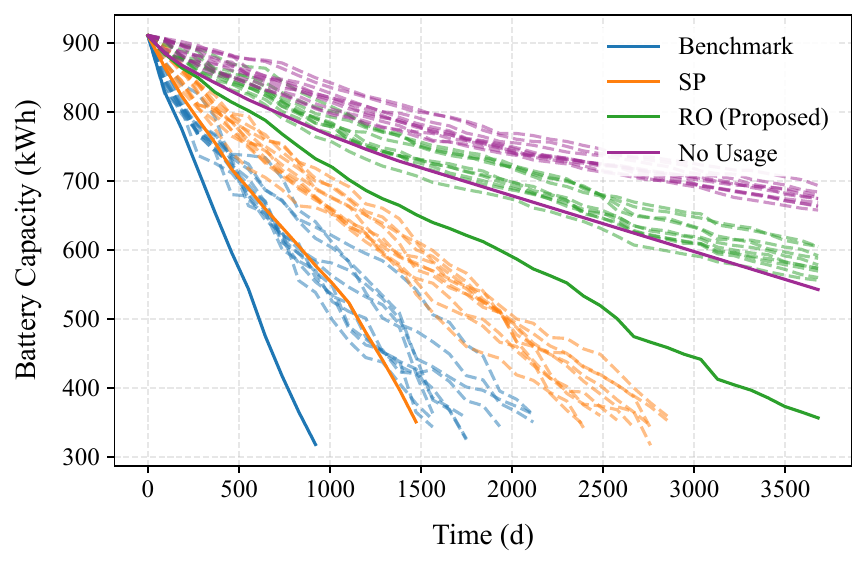}
\caption{Battery capacity degradation curves of different methods. The solid lines represent the 90\% worst-case degradation path, while the dashed lines show 10 Monte Carlo simulated paths under the same policy.}
\label{fig:degradation}
\end{figure}

These methods are used to obtain the value of $\boldsymbol{\theta}$. Once obtained, life-cycle simulations are conducted either for the worst-case scenarios or by Monte Carlo simulation. The results are summarized in Figure~\ref{fig:degradation} and Table~\ref{tab:comparison}. Figure~\ref{fig:degradation} illustrates the difference in battery life. The benchmark method aggressively leverages the battery to decrease operational costs, causing rapid degradation and the shortest lifespan. The SP method extends the battery life by accounting for degradation, but the proposed RO method achieves the longest battery life by adopting a more conservative dispatch strategy.

\begin{table}[H]
\renewcommand{\arraystretch}{1.2}
\caption{Method Comparison}
\label{tab:comparison}
\centering
\footnotesize
\vspace{0.5em}
\begin{tabular}{lcccc}
\hline
& Benchmark & SP & RO & No Usage \\ 
\hline
$\theta^{EFC}$ (\$/kWh) & $0.000$ & $0.045$ & $0.087$ & - \\
$\theta^{DOD}$ (\$/kWh) & $0.000$ & $0.063$ & $0.066$ & - \\
$\theta^C$ (\$/kW) & $0.000$ & $0.019$ & $0.073$ & - \\
$\theta^D$ (\$/kW) & $0.000$ & $0.041$ & $0.056$ & - \\
90\%-Worst Cost (\$$10^6$) & $1.304$ & $1.168$ & $\mathbf{1.142}$ & $1.204$ \\
95\%-Worst Cost (\$$10^6$) & $1.380$ & $1.226$ & $\mathbf{1.178}$ & $1.204$ \\
Mean Battery Life (d) & $1,665$ & $2,668$ & $\mathbf{3,680}$ & $\mathbf{3,680}$ \\
\hline
\end{tabular}
\end{table}

Table~\ref{tab:comparison} quantifies the behaviors under different methods. The optimized $\boldsymbol{\theta}$ parameters for the RO method are higher than the SP method, reflecting that RO assigns a higher penalty to degradation to hedge against uncertainty. The proposed RO method achieves the lowest total life-cycle cost in both the 90\% and 95\% worst-case scenarios, demonstrating its ability to deliver a more robust performance over the long term. At the same time, the proposed method has the longest mean battery life, showing that the battery life can be substantially extended by applying a higher degradation penalty in the dispatch. If the ESS is not in use during microgrid operation, its capacity decreases due to calendar aging, and the corresponding total life-cycle cost is higher than the proposed method.

\subsubsection{Parameter Sensitivity Analysis}

We now analyze how the optimal strategy and life-cycle costs are affected by the quantile setting, temperature, and initial ESS capacity.

As shown in Table~\ref{tab:quantile}, increasing $q$ makes the strategy more conservative, since it optimizes against a more pessimistic worst-case scenario. The mean battery life does not increase monotonically, revealing a complex trade-off between operational costs and battery degradation.

\begin{table}[H]
\renewcommand{\arraystretch}{1.2}
\caption{Results for Different Quantiles}
\label{tab:quantile}
\centering
\small
\vspace{0.5em}
\begin{tabular}{lcccc}
\hline
Quantile & $0.80$ & $0.85$ & $0.90$ & $0.95$ \\ 
\hline
Objective (\$$10^6$) & $1.076$ & $1.119$ & $1.142$ & $1.171$ \\
Mean Battery Life (d) & $2,916$ & $3,459$ & $3,680$ & $2,953$ \\
\hline
\end{tabular}
\end{table}

As predicted by our degradation model, ambient temperature has a significant impact on life-cycle cost in Table~\ref{tab:temperature}. Both low (5°C) and high (50°C) temperatures are stressful environments that accelerate degradation, leading to higher total costs compared to the base case of 35°C. The effect is particularly strong at 50°C, where the significantly accelerated degradation reduces the mean battery life by nearly three years.

\begin{table}[H]
\renewcommand{\arraystretch}{1.2}
\caption{Results for Different Ambient Temperatures}
\label{tab:temperature}
\centering
\small
\vspace{0.5em}
\begin{tabular}{lccc}
\hline
Temperature (°C) & $5$ & $35$ & $50$ \\ 
\hline
Objective (\$$10^6$) & $1.173$ & $1.142$ & $1.177$ \\
Mean Battery Life (d) & $3,680$ & $3,680$ & $2,603$ \\
\hline
\end{tabular}
\end{table}

Table~\ref{tab:capacity} shows the results for different initial ESS capacities, with a fixed investment cost. A larger ESS provides greater operational flexibility, allowing the microgrid to meet its needs with less stress on the battery, such as lower DOD values. This flexibility leads to lower operational costs and thus reduces the total life-cycle cost.

\begin{table}[H]
\renewcommand{\arraystretch}{1.2}
\caption{Results for Different Installed ESS Capacities}
\label{tab:capacity}
\centering
\small
\vspace{0.5em}
\begin{tabular}{lcccc}
\hline
Installed Capacity & \multirow{2}{*}{$683.1$} & \multirow{2}{*}{$910.8$} & \multirow{2}{*}{$1138.5$} & \multirow{2}{*}{$1366.2$} \\ 
of Battery (kWh) & & & & \\
\hline
Objective (\$$10^6$) & $1.199$ & $1.142$ & $1.047$ & $0.958$ \\
Mean Battery Life (d) & $3,680$ & $3,680$ & $3,468$ & $3,533$ \\
\hline
\end{tabular}
\end{table}

\section{Conclusion}

This paper develops an XGBoost-based probabilistic battery degradation model using experimental data, and proposes a robust parametric dispatch framework considering the DDU of battery degradation. The proposed method optimizes the trade-off in the dispatch model between immediate operational costs and long-term battery degradation using life-cycle simulation. In addition, it manages to model the dynamic and bidirectional relationship between dispatch decisions and battery degradation, utilizing the RO framework with DDU. Case study results demonstrate the performance of the battery degradation model, and show that the proposed dispatch method is more robust than conventional approaches. 

Future work could extend the proposed framework in the following directions: 1) Training and validating the degradation model on more datasets that include diverse battery types. 2) Developing a multi-scale degradation model that bridges cell-level behavior with pack-level dynamics, and accounts for thermal management strategies. 3) Integrating the initial ESS sizing optimization into the framework to achieve a more holistic life-cycle cost reduction for the microgrid.

\section{Acknowledgments}

This work was supported by the State Key Laboratory of Environmental Adaptability for Industrial Products Open Research Fund (Grant No. 2025EASKJ-007).

\end{multicols}

\end{document}